# Probing the Interior Environment of Carbon Nano-test-tubes

Andrew A. R. Watt, Mark R. Sambrook, Stanislav V. Burlakov, Kyriakos Porfyrakis, G. Andrew D. Briggs

Department of Materials, Oxford University, Parks Rd, Oxford, OX1 3PH, Fax: +44 1865 273725, E-mail: andrew.watt@materials.ox.ac.uk

**ABSTRACT**

We report the filling of single walled carbon nanotubes with an electron spin-active fullerene species where a nitroxide radical is functionalized on the fullerene cage. High resolution transmission electron microscopy (HRTEM), optical absorption and electron spin resonance (ESR) are used to determine the rotational behavior of the encapsulated molecules and determine the polar nature of the nanotube interior.

## 1. Introduction

The insertion and assembly of small molecules inside carbon nanotubes has been demonstrated via a number of methods [1-6]. Carbon nanotubes can be used to define one dimensional chains of molecules with useful magnetic, optical or electronic properties for applications such as quantum information processing [7,8]. Currently, research is focused upon understanding the modification and interaction of molecules confined inside single walled carbon nanotubes [9]. For example Yanagi *et al* have optically measured the modification of electronic properties of encapsulated ß-carotene molecules [10], Koshino *et al* observed the rotation of individual $C_{12}$ and $C_{22}$ molecules [11], and Simon

*et al* measured the electron spin modification of N@$C_{60}$ endohedral fullerenes. The last system is a 1D spin chain where single nitrogen atoms are localized within and shielded by a $C_{60}$ cage preventing strong interaction with the nanotube[1,2]. In this paper we exohedrally functionalize a fullerene with a nitroxide radical which has a strong electronic interaction with its surrounding environment and use ESR to probe the interior environment of the nanotube.

**2. Method**

Single walled nanotubes (SWNT) produced by arc-discharge where purchased from Aldrich and purified via the following steps **1.** Treatment in a kitchen microwave for 5 minutes at low power **2.** Heating in air at 325 $^o$C for 1 hour. **3.** Refluxing in 37% hydrochloric acid for 2 hours. **4.** Filtering and washing with de-ionized water. **5.** Refluxing in hydrogen peroxide for 2 hours, **6.** Filtering and washing with de-ionized water. **7.** Drying at 200 $^o$C and storage in a nitrogen dry-box.

The $C_{60}$-TEMPO conjugate was prepared according to literature procedures [13] and characterized using MALDI mass spectrometry UV-VIS absorption, ESR and FTIR (see supporting information). Nanotubes where filled with the functionalized fullerenes by taking 3 mg of SWNTs and 1 mg of TEMPO-$C_{60}$ which were dehydrated and cycled three times through the supercritical phase of carbon dioxide over a 72 hour period using a Quorum CPD7501. The resulting product mixture was dispersed in toluene, ultrasonicated for 2 hours, centrifuged and the supernatant removed. This was repeated several times with toluene, carbon disulfide and methanol until no further colour was observed in the supernatant thus ensuring complete removal of unfilled fullerene species from the sample [14]. The resulting carbon material was then filtered and washed with toluene. Samples for high resolution transmission electron microscopy (HRTEM) were prepared by dropping dilute dispersions of the peapods ultrasonicated in dichloroethane onto lacey amorphous carbon supports. These were analyzed using a JEOL JEM-4000EX high resolution electron microscope with an information limit of 0.12nm operating at an accelerating voltage of 100 kV equipped with a Gatan Digital Camera. Samples for absorption and ESR were prepared by ultrasonication in a butylamine/THF (1:1) solution; this formed a stable colloidal solution, great care was taken to ensure peapod and control samples where subjected to the same treatments. Absorption measurements where made using a Jasco UV-VIS-NIR V-570 spectrometer and ESR was performed using a Magnetech Miniscope MS200 equipped with a liquid nitrogen bath cryostat, hyperfine and g-factor where fitted using Easyspin [15].

## 3. Results and Discussion

The HRTEM image in figure 1 (b) shows that TEMPO-$C_{60}$@SWNT peapods were successfully prepared using the supercritical method. We estimate from HRTEM that in an average sample at least 20% of the nanotubes are completely filled. The TEMPO moiety is not observable due to low contrast. Nanotube width and molecular separation within the nanotube can be accurately determined by taking a two dimensional Fast Fourier Transform of the real space peapod image as shown in figure 1(c) [16]. We determine that the nanotube is 1.45 ± 0.05 nm wide and the fullerene spacing is 1.01 ± 0.05 nm. The fullerenes are as closely packed as those observed in $C_{60}$ which is expected for a fullerene functionalized with a short [17] Figure 2 compares the UV-VIS-NIR absorption spectra of purified SWNTs and TEMPO-$C_{60}$@SWNTs. Encapsulation modifies the absorption spectra considerably, the metallic and semiconducting transitions observed in the nanotube at 690 nm and 1022 nm are red-shifted to 757 nm and 1070 nm, respectively. This corresponds to the shift observed by Bandow et. al. for plain $C_{60}$ peapods [18]. We postulate that this is due to modification of the nanotube density of states by the encapsulated fullerene species [19]. The change in absorption is also good evidence that a good proportion of nanotubes are filled. There is no appreciable absorption signature from the TEMPO-$C_{60}$ in the wavelength range shown in figure 2.

Room temperature ESR of SWNT, TEMPO-$C_{60}$ and TEMPO-$C_{60}$@SWNT peapods are shown in figure 3. Initial observations reveal a signal from the carbon nanotube in both empty and filled carbon nanotubes and the survival of the TEMPO radical after the filling process. Significantly higher microwave power was required in order to observe an ESR signal from the encapsulated TEMPO-$C_{60}$ species, indicating the absorption of microwaves by the carbon nanotube and shielding of the nitroxide radical. Perturbations are also observed in the hyperfine splitting, g-factor and relative peak intensities of the TEMPO ESR signal upon encapsulation in carbon nanotubes. The origin of the carbon nanotube ESR signal is a matter of debate, it could be due to nanotube conduction electrons [20], nanotube defects [21], amorphous carbon [22], graphitic carbon [23], or some residual catalyst impurity.

The magnitude of the hyperfine coupling in nitroxide radicals is known to vary with environment; increasing polarity and hydrogen-bond donor ability of a solvent result in larger hyperfine splittings [24]. The hyperfine splitting for TEMPO-$C_{60}$@SWNT was 15.0 Gauss and TEMPO-$C_{60}$ was 14.5 Gauss this indicates that the environment within the nanotube is non-polar. A change in the g-factor of the nitroxide radical is also observed upon encapsulation in the carbon nanotube from 2.0045 to 2.0082. This is analogous to that observed for TEMPO molecules dispersed in liquid crystals upon a phase change from isotropic to nematic and indicates that the radical is aligned within the carbon nanotube in the colloidal dispersion [25].

Although characteristic of nitroxide free radicals, the ESR signal of the TEMPO-$C_{60}$@SWNT peapod reveals both broader linewidths and a modulation in the relative intensities of the three signals. In non-viscous solvents nearly all nitroxides exhibit three equally spaced sharp lines as a result of rapid isotropic tumbling averaging anisotropic effects. Slowing rotational motion gives rise to incomplete averaging and unequal broadening of the three peaks and so the spectrum becomes asymmetric [26].

Although nanotubes possess rotational freedom in colloidal solution, any tumbling or rotations perpendicular to the length of the nanotube will be more restricted than the rotational freedom of the free TEMPO-$C_{60}$ molecule itself. If an x-axis is defined as the N-O bond of the nitroxide radical then the observed spectral modification can arise if this rotation remains free and the perpendicular z- and y-axes are restricted. This situation can only arise in this system if the N-O, or x-axis, is aligned with the walls of the carbon nanotube. This observation is in agreement with the effects observed when TEMPO-based molcules are aligned in liquid crystal matrices [27].

ESR measurements at 77 K of single-walled carbon nanotubes, TEMPO-$C_{60}$ and TEMPO-$C_{60}$@SWNT peapods are shown in figure 4. The hyperfine splitting in both cases is asymmetric with TEMPO-$C_{60}$@SWNT having 18 and 22 Gauss and TEMPO-$C_{60}$ 22 and 32 Gauss In the low temperature case there is molecular rotation restriction for both solvent and nanotube matrices indicated by the hyperfine asymmetry [23]. However the difference in hyperfine is the converse of the room temperature measurement where the peapod sample exhibited a larger splitting. This shows that inside the nanotube the TEMPO-$C_{60}$ is in a less rotationally rigid environment at low temperature due to minimal interaction with frozen solids. A change in the g-factor of the nitroxide radical is also observed upon encapsulation in the carbon nanotube from 2.007 to 2.005. Attempts to fit the hyperfine and g- factor where hampered by the strong background signal in the composite, making it hard to compare line shape, relative position and peak intensity asymmetry (supporting information).

## 4. Conclusion

In conclusion it has been demonstrated that nitroxide based radicals remain stable when encapsulated in single walled carbon nanotubes and that the ESR signal can be readily observed. The carbon nanotube structure provides a high degree of shielding from microwave radiation, is relatively non-polar and its rotational properties can be observed by modification of the ESR intensities of the encapsulated nitroxide radical. At low temperature the radical has a greater degree of rotation inside the nanotube compared to a frozen solid.

**Acknowledgments**

This research is part of the QIP IRC http://www/qipirc.org (GR/S82176/01). GADB is supported by an EPSRC Professorial Research Fellowship (GR/S15808/01). We thank the EPSRC National Mass Spectrometry Service Centre, University of Wales, Swansea, for sample characterization.

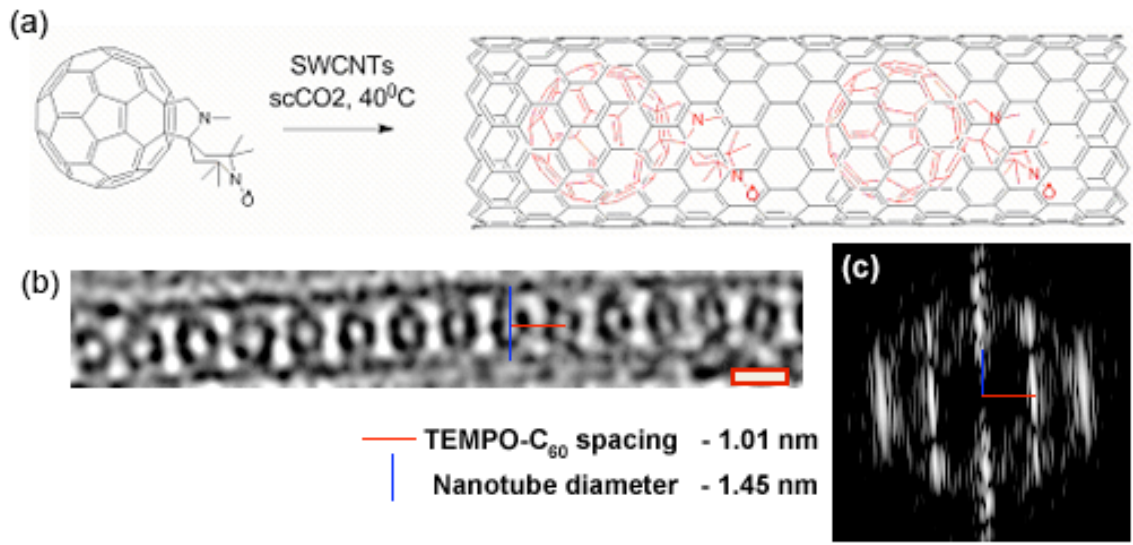

Figure 1 (a) Schematic of filling of single-walled carbon nanotubes with TEMPO-$C_{60}$@SWNT using supercritical carbon dioxide. (b) High resolution transmission electron micrograph of the TEMPO-$C_{60}$@SWNT peapod. (c) 2D-FFT of real space image showing nanotube diameter and fullerene spacing.

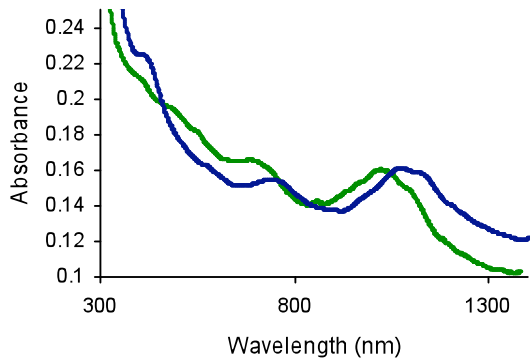

Figure 2. UV/visible absorption spectra of carbon nanotubes (**green**) and TEMPO-C$_{60}$@SWNT peapods (**blue**).

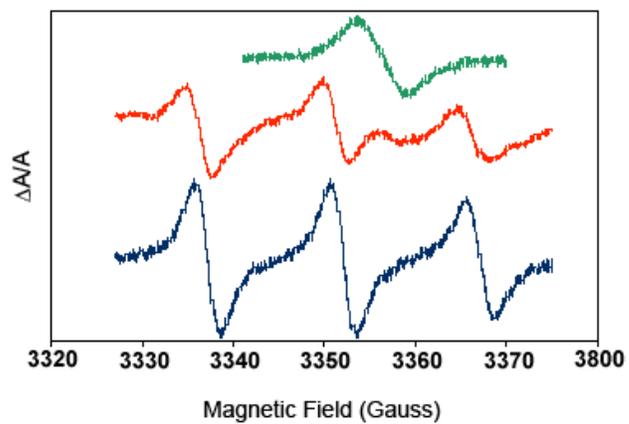

Figure 3. 300 K electron spin resonance spectra of single-walled carbon nanotubes (**green**), TEMPO-$C_{60}$ (**red**) and TEMPO-$C_{60}$@SWNT peapods (**blue**).

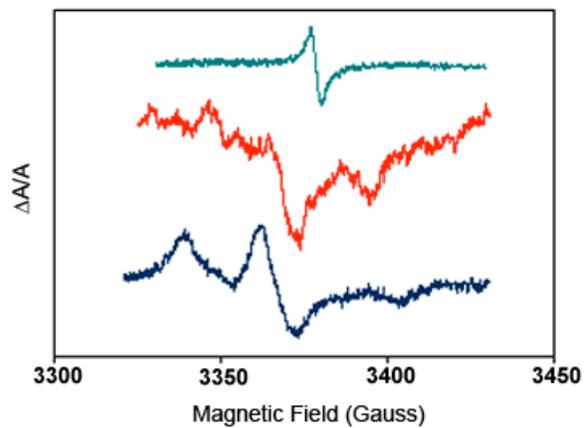

Figure 4. 77 K electron spin resonance spectra of single-walled carbon nanotubes (**green**), TEMPO-$C_{60}$ (**red**) and TEMPO-$C_{60}$@SWNT peapods (**blue**).